# Modeling a fluid to solid phase transition
# in snow weak-layers. Application to slab avalanche release

François Louchet

**Abstract:** Snow slab avalanche release usually results from failure of weak layers made of loose ice crystals. In previous field experiments, we evidenced for the first time an interesting stress-driven transition in the weak layer between a granular fluid and a solid phase. We propose here an original model involving the kinetics of ice grains bonds failure and reconstruction. The model evidences a sudden transition between two drastically different types of weak layer behaviors. It accounts for the characteristics of both the studied fluid-solid transition and for slab avalanche release observations. It may possibly apply to a number of other granular materials.

**Keywords**: granular media; snow slab avalanches; avalanche release; weak layers; basal crack; slurry; fluid-solid transition; shear strain rate; whumpf; settlement.

## 1. Introduction

Snow slab avalanches release usually results from initial failure of a low density weak layer (*WL*) that often separates the upper slab from the older snow substrate. Such a failure, that may or may not result in avalanche release, usually propagates below the slab over distances ranging from meters to kilometers, and is often associated with a downward displacement (or "settlement") of the slab, and an audible "whumpf", an onomatopoeic term for the muffled noise produced by the settlement.

Weak layers are most often made of layers of faceted crystals grown underneath snow slabs, or of surface hoar that has been buried under recent snowfalls. They consist of granular aggregates of low density polyhedral ice grains bonded by brittle ice bridges. They are recognized to play a key role in snow avalanche triggering processes, though the precise underlying mechanisms are still under strong debate. It is therefore of interest to understand the details of the *WL* behavior, in order to be able to predict which conditions may favor avalanching instead of simple "whumpfing".

The present work is based on field experiments carried out a few years ago in the Orelle ski resort (french Alps) [1] in order to characterize more precisely the *WL* behavior and its possible consequences on avalanche release processes. Shortly after being collected from the *WL* with a shovel, the material was observed to behave as a granular slurry, hereafter named fluid (*F*), made of grains flowing like dry rice, suggesting that a mechanically disturbed *WL* may possibly act as an easy glide surface for the overlying slab. However, when left undisturbed for a few seconds, the fluid clotted into a solid (*S*). When mechanically disturbed again (e.g. by a mechanical shock), the *S* phase turned back to the *F* one if the disturbance was large enough.

In the present paper, this transition is explored using a mean-field model, in the spirit of the theory of dynamical systems, analyzing the kinetics of bond failure and reconstruction. We discuss the possible consequences on snow avalanche release.

Possible extensions of the model to some other phenomena involving granular media, particularly in geophysics, are also contemplated.

## 2.General model

### 2.1. Modeling

We describe the fluid medium $F$ as a slurry made of solid clusters (aggregates of clotted grains) embedded in a granular "liquid" $L$ (free grains). We compute the rate at which solid clusters form, grow or disaggregate during shear displacements that control contact times for both grain-cluster and grain-grain interactions. The concentrations of grains belonging respectively to clusters, granular liquid and cluster/granular-liquid interfaces are labelled $N_C$, $N_L$ and $N_i$, in such a way that:

$$N_L+N_c+N_i = 1 \tag{1}$$

The system is in a pure solid state when $N_c=1$, and in a pure "liquid" state when $N_c=0$. Topologically, in between those two extreme situations, the system can in principle be found in three different states: i) <u>fluid phase</u> (percolating "liquid" $L$ that may contain solid clusters), ii) <u>bipercolated state</u>, in which both "liquid" and solid phases percolate through the system, and iii) <u>"pseudo-solid" state</u>, i.e. percolating solid phase that may contain "liquid" bubbles.

Flow in the two last states is controlled by creep of the percolating solid phase, which is of quite a different nature, with a flow rate orders of magnitude slower than that of the fluid state. For this reason, we neglect the creep rate of states ii) and iii), that will be considered to behave as a purely elastic solid, and investigate the behavior of the system starting from the fluid (i.e. slurry) state.

The solid phase is expected to percolate through the system at a well defined threshold $N_p$, around 0.3 for randomly packed spheres [2,3]. Due to shear flow, this threshold is expected to be somewhat larger in dynamical conditions than in static ones. Nevertheless, we shall restrict the validity of the present model to $N_c$ values smaller than 0.3.

In this domain, and due to minimization of interface energy, clusters should be more or less spherical, with an average radius $R$. Also assuming spherical grains of radius $r$:

$$N_i \approx 4\pi R^2/\pi r^2 = 4R^2/r^2 \quad , \quad N_c \approx (4/3)\pi R^3/(4/3)\pi r^3 = R^3/r^3$$

and therefore $N_i$ should be related to $N_c$ by:

$$N_i \approx 4 \, N_c^{2/3} \tag{2}$$

We first study the system loaded in imposed strain rate conditions. The "liquid" phase is fed by particles taken from cluster interfaces; this process is driven by the shear of the fluid phase, and its kinetics therefore assumed to be proportional to shear strain rate $\dot{\gamma}$ and to grain concentration at interfaces $N_i$.

On the other hand, particles are taken out from the "liquid" phase $L$ as they form bonds with either other "liquid" particles (cluster nucleation) or interface grains (cluster growth). Regarding bond formation between two grains in $L$, it is worth noting that, in contrast with gas kinetics in which the reaction rate between molecules scales as the square of their concentration, we deal here with a dense phase in which each grain of $L$ is always in contact with other ones. As a consequence, the reaction rate is not limited by the probability of grain collisions in a dilute medium, but scales as the grain concentration $N_L$ only (actually $N_L / 2$ in order to avoid counting twice the same grain). For the same reason, the reaction rate between one grain in $L$ and a cluster interface scales as the concentration $N_i$ of interface grains only. Finally, the reaction kinetics of such bond formation is taken proportional to the contact time between "liquid" and "interface" grains, that obviously scales as $\left(1/\dot{\gamma}\right)$.

As a consequence, the global reaction kinetics describing the evolution rate $dN_C/dt$ of the concentration of grains belonging to clusters (hereafter labelled $\dot{N}_C$) as a function of $N_C$, is given by the evolution equation:

$$\dot{N}_C = -A\dot{\gamma}\,N_i + \frac{B}{\dot{\gamma}}\left[\frac{N_L}{2} + N_i\right] \tag{3}$$

Or, using eqs. (1) and (2):

$$\dot{N}_C = -4A\dot{\gamma}\,N_C^{2/3} + \frac{B}{\dot{\gamma}}\left[\frac{1-N_C}{2} + 2N_C^{2/3}\right] \tag{4}$$

where $A$ is a dimensionless constant, and $B=B(T)$ (with dimension $t^{-2}$) is a function of temperature, that characterizes diffusion kinetics of water molecules at cluster interfaces:

$$B(T) \propto \exp\left(\frac{-Q}{kT}\right) \tag{5}$$

$Q$ being the activation energy for the diffusion process, and $k$ the Boltzmann constant.

Actually, it is easier to argue in terms of imposed stress (determined by both slope angle and slab weight) than of imposed strain rate. In order to transform equation (4) into an imposed-stress equation, we need to define a viscosity $\eta$ characteristic of the fluid phase only:

$$\eta = \tau / \dot{\gamma} \tag{6}$$

where $\tau$ is the shear stress.

Defining a physically based viscosity of such a diphasic medium is not straightforward, as it has to take into account particle collisions, energy dissipation, etc. The viscosity of a suspension of spheres in a liquid medium, derived on such bases by Einstein in 1906 and 1911, was analyzed in [4], and can be expressed in our notations as:

$$\eta = \eta_0 \frac{1+N_C/2}{1-2N_C} \tag{7}$$

where $\eta_0$ is the residual viscosity of the pure "liquid" (i.e. without any solid cluster). This expression is valid only for dilute suspensions, i.e. $N_c$ close to 0, and diverges for $N_C = 1/2$. The validity of such a viscosity for increasing $N_C$ values is also limited by topological constraints due to solid percolation, as mentioned above. In contrast with all previous models, a percolation-based approach of the viscosity of concentrated suspensions was proposed in [5]. However, for the sake of simplicity, and because we are only interested in the evolution of the fluid phase, we shall use here Einstein's expression, which is a good approximation for $N_c$ values smaller than the percolation threshold $N_p \approx 0.3$.

Using Eqs. (6) and (7), Eq. (4) writes:

$$\dot{N}_C = -4A \frac{\tau}{\eta_0} N_C^{2/3} \frac{1-2N_C}{1+N_C/2} + B \frac{\eta_0}{\tau} \frac{1+N_C/2}{1-2N_C} \left[ \frac{1-N_C}{2} + 2N_C \right] \tag{8}$$

We are only interested in the time evolution of $N_C$, i.e. the sign of $\dot{N}_C$, which is the same as that of the dimensionless variable $\psi$ defined by:

$$\psi = \dot{N}_C \frac{1}{B} \frac{\tau}{\eta_0} \tag{9}$$

$$= -4 \frac{A}{B} \left( \frac{\tau}{\eta_0} \right)^2 \dot{N}_C^{2/3} \frac{1-2N_C}{1+N_C/2} + \frac{1+N_C/2}{1-2N_C} \left[ \frac{1-N_C}{2} + 2N_C^{2/3} \right]$$

$$= N_C^{2/3} \left[ 2 \frac{1+N_C/2}{1-2N_C} - \Sigma \frac{1-2N_C}{1+N_C/2} \right] + \frac{1-N_C}{2} \frac{1+N_C/2}{1-2N_C} \tag{10}$$

where the influence of stress is represented by the dimensionless variable :

$$\Sigma = 4 \frac{A}{B} \left( \frac{\tau}{\eta_0} \right)^2 \tag{11}$$

Fig. 1 shows typical $\psi(N_C)$ curves parametrized by $\Sigma$ for $N_c$ values between 0 and 0.3, as justified above.

For $\Sigma < \Sigma^*$, i.e. low stresses and (or) large viscosities (top curve), $\psi(N_C)$ is always positive, and so is $dN_C/dt$. Starting from any initial $N_C$ value, $N_C$ continously increases and the fluid eventually clots into the solid phase.

By contrast, for $\Sigma > \Sigma^*$, i.e. large stresses and (or) small viscosities (bottom curve), the curve exhibits a negative minimum and intersects the $N_C$ axis at $A$ and $R$, that are the fixed points of the system ($dN_C/dt=0$), corresponding to two $N_C$ values, $N_R$ and $N_A$ ($N_R > N_A$). $A$ is an attractor, since an upward fluctuation of $N_C$ results in a negative $dN_C/dt$ value, that brings the $N_C$ value back to $N_A$, and conversely for a downward fluctuation. A similar argument shows that $R$ is a repulsor. As a consequence, starting from any $N_C$ value larger than $N_R$ drives the system towards the solid phase, whereas starting from any $N_C < N_R$ brings the system to the attractor $A$.

In between, both fixed points merge for $\Sigma = \Sigma^*=7.04$ (intermediate curve), corresponding to $Nc=N_c^*\approx 0.07$, which means that the stable fluid found at $N_A < N_c^*$ contains more than 93% "liquid". This proportion increases even further with $\Sigma$, consistently with the assumption, used in equation (7), that we are in the validity domain of Einstein's viscosity expression.

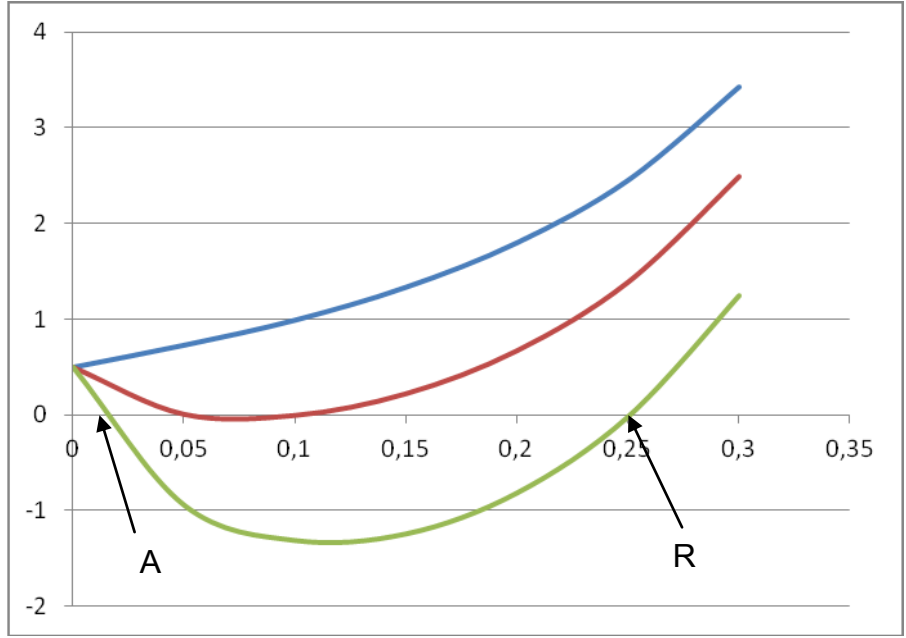

**Fig.1: typical ψ(N_C) curves parametrized by Σ:** *top: Σ=1 ( low stresses), middle: Σ= Σ\*= 7,04 (transition), bottom: Σ=15 (high stresses). A and R are respectively the attractor and the repulsor of the system.*

## 2.2. Comparison with previous field observations

Such results can now be qualitatively compared to the field observations recalled in the introduction. Let us start from a granular fluid already flowing on a slope under its own weight, like dry rice. If the slope is steep enough, the driving force (gravitational shear stress) is large, and $Σ$ may exceed the threshold value $Σ^*$. In this case, the system is trapped around the attractor $A$, and the fluid goes on flowing. If the slope is gradually reduced, so is the shear stress, $Σ$ decreases, the attractor shifts towards slightly larger $N_C$ values, but the fluid keeps flowing until $Σ$ reaches the $Σ^*$ value for which the two fixed points $A$ and $R$ merge. Beyond this point, $ψ$ becomes positive everywhere, and the fluid clots into the solid phase $S$, as observed.

Let us now start from an unloaded clotted solid ($Σ=0$, $N_C= 1$). We first load the system in shear, up to a stress corresponding to a non-zero $Σ$ value. Since the system is solid, there is no flow, $\dot{γ}=0$, and it cannot depart from the solid state, whatever the stress might be. However, keeping the same $Σ$ value, we can change the initial conditions giving a mechanical shock in order to break bonds and temporarily decrease the $N_C$ value and bring the system into the fluid domain, i.e. $N_C < N_p ≈ 0.3$. If $Σ$ is smaller than $Σ^*$ (moderately slanted slope, top curve in Fig. 1), $ψ$ being always positive, the system clots back again. On the opposite, for $Σ$ values larger than $Σ^*$ (steep slope, bottom curve in Fig. 1), two situations may be contemplated. A weak shock may temporarily bring $N_C$ to a value comprised between $N_R$ and $N_p$. In this case, despite the large $Σ$ value, $ψ$ remains positive, and the system clots back again. By contrast, a stronger shock may bring $N_C$ to a value smaller than $N_R$, i.e. into the attraction basin of $A$. In this case, the system readily becomes fluid, converges at $N=N_A$, and starts flowing down. For very large $\Sigma$ values, $N_R$ being large, even a tiny shock may destabilize the system. All these predictions qualitatively agree with observations.

## 3. Application to snow slab avalanche release

### 3.1. Assumptions and definitions

The present model has an obvious application to snow avalanche release, regarding basal crack nucleation and propagation at interfaces between slabs and substrates. It is now widely acknowledged that slab avalanche release does not result from the additional weight of a skier for instance, supposed to help the slab weight to exceed the weak layer friction. The skier's weight is indeed negligible in comparison of the weight of the involved slab (typically several thousands tons). Instead, local and short loadings on slabs, due to skiing or snowboarding or to artificial triggering explosive devices, may be responsible for the initiation of a so-called basal crack that may further expand due to the weight of the slab itself (Fig. 2).

Previous simplified models considered the stability of such basal cracks under shear loading only, essentially based on Griffith's concepts [6-8] or on more complicated but equivalent ones [9-11]. However, the weak layer is a non-compact medium that may easily collapse under loads having a compressive component, as already confirmed some time ago [12-13] and clearly shown recently using Propagation Saw Tests (PST) [14-16]. Basal cracks are therefore initiated by a local combination of weak layer collapse and shear failure modes, varying between the two limiting cases of pure shear and pure collapse [17]. It can be inferred from the field observations mentioned in the introduction [1] that such a combined collapse-shear failure of the *WL* results in a fluid phase layer prone to downslope glide. The question is whether the fluid phase *F* will remain fluid and allow avalanching, or readily clot into a solid phase *S*, that would stop the triggering process. This question is now discussed in terms of the above model.

In the present analysis of triggering mechanisms, we shall use the following definitions:
$h$: "real" slab thickness, i.e. measured perpendicular to the slab
$h_{//}$: slab thickness measured vertically
$w$: thickness of the collapsible part of the *WL* measured perpendicular to the slab
$w_{//}$: thickness of the collapsible part of the *WL* measured vertically
$\delta$: residual weak layer thickness after collapse, measured perpendicular to the slab
$w+\delta$: weak layer thickness before collapse, measured perpendicular to the slab

For the sake of simplicity, we first consider an <u>infinite slope</u>, with a slope angle $\alpha$ with respect to horizontal. Assuming no wind conditions, we consider that snow falls vertically. As a consequence, the resulting slab has a constant vertical thickness $h_{//}$, corresponding to a "real" thickness $h=h_{//} \cos \alpha$ measured perpendicular to the slope, that obviously decreases at constant $h_{//}$ for increasing slope angles.

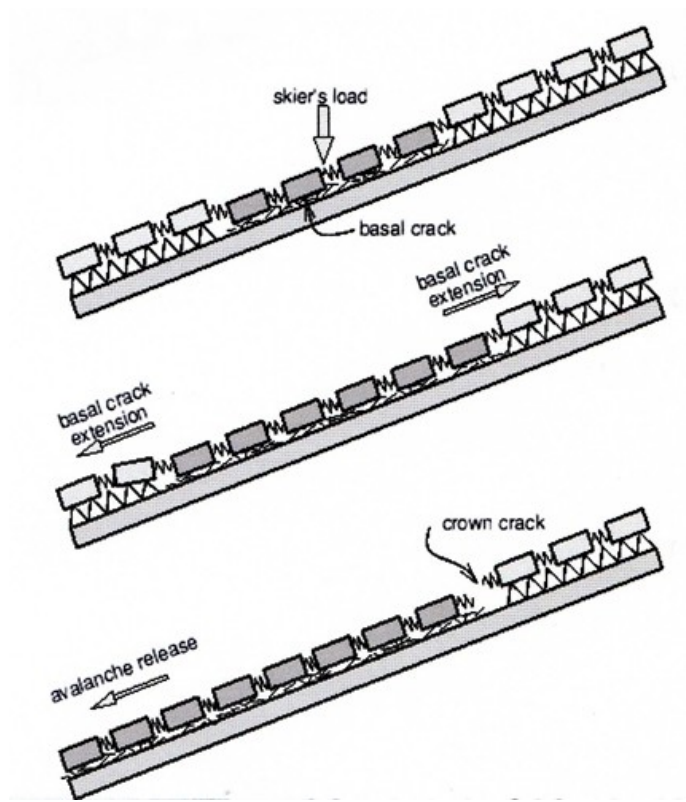

**Fig. 2: Schematic of a slab avalanche triggering process:** *Top: the WL locally fails (e.g. under a skier's impulse). Middle: the crushed area (basal crack) starts expanding in all directions driven by slab weight. Bottom: due to the slope-parallel weight component of the suspended part of the slab, a crown crack opens at the top, allowing large scale avalanching.*

*WL*s usually consist of either faceted grains or buried surface hoar. Faceted grains formation within the *WL* is a thermodynamical process driven by temperature gradients, that does not directly depend on gravity. In a same way, surface hoar formation, that occurs during humid and clear nights, results from crystal growth at snow surface driven by humidity and temperature gradients, independently of the thickness of the snow layer on which it grows. We can therefore consider as a first approximation that, in both cases of faceted crystals or of buried surface hoar, *WL* thicknesses $w+\delta$ and $w$ measured perpendicular to the slab, as well as *WL* mechanical properties, are constant along the slope. Avalanche release results from two successive stages, that will be examined hereafter.

3.2. *WL* failure initiation

This first stage takes place under a combination of compression and shear load components. A local *WL* failure can be initiated for instance by a skier, whose impulse may locally crush the *WL* through a deformation of the slab and result in a reduction of *WL* thickness from $w+\delta$ down to $\delta$. As mentioned in the introduction, the crushed *WL* material just after failure is in a fluid state *F*. Once initiated, the failed zone may extend under a "driving force" equal to the work of the slab weight per square meter $\rho g h$ along its vertical displacement $w_{//}$. From above definitions, this work can be written:

$$\rho g h w_{//} = \rho g (h_{//} \cos\alpha)(w/\cos\alpha) = \rho g h_{//} w \qquad (12)$$

Under the above assumption that both $h_{//}$ and $w$ are independent of the slope angle, so should be the driving force for *WL* failure.

The critical radius of the collapsing zone (in Griffith's sense [6]), above which it may extend spontaneously, can be easily computed as a balance between the driving force on the one hand, and the resistant force opposing crack extension, determined by the energy required for crushing the *WL*, on the other hand.

Such a driving force for crushing the *WL* was calculated under the assumption of a constant $h$ (and not $h_{//}$) [17], which might lead to fairly different conclusions than ours. However, both their driving force and ours coïncide on flat terrain, where $h = h_{//}$. Our remark above, that the driving force in the case of vertical snowfall is independent of slope angle (Eq. (12)), allows us to generalize to all kinds of slopes their results on flat terrain.

One of their results was that the critical size is fairly low, with a typical value on flat terrain of a few decimeters. This result is valid in our case for all slope angles. This means that a small local collapse (larger than a few decimeters) is likely to readily propagate along the whole *WL*, whatever the slope.

### 3.3. Subsequent *WL* shearing and downslope slide of the slab

The question is now whether this crushed zone would result in avalanche release or not. Since the *WL* is already collapsed, this second stage is driven by the shear component of the load only. Two different situations may be contemplated, depending on whether the shear stress $\tau$ in the *WL* is smaller or larger than the critical shear stress value $\tau^* = \eta_o \sqrt{B\Sigma^*/4A}$ ) defined using Eq. (11), and corresponding to the value $N_C = N_C^*$.

i) for sufficiently large slope angles and slab weights, the shear stress $\tau$ in the *WL* is larger than the critical shear stress $\tau^*$. In this case, the strain rate is sufficiently large to maintain the *WL* in the fluid state, and the slab can slide down.

ii) for small slope angles and (or) slab weights, the conditions are such that the shear stress is less than $\tau^*$, and the *WL* clots into the *S* phase after a time $\Delta t$. However, during the clotting process, the collapsed zone continues to expand, leaving a fluid zone in its wake. Assuming that the collapsed zone has a circular shape, the collapsed area consists of a central disk of already clotted snow surrounded by a ribbon of still fluid granular material having almost no shear resistance.

Let $V$ be the propagation velocity of the border of the collapsed zone, that has been schematized by the propagation of a solitary wave and theoretically computed in [18]. As clotting starts at a time $\Delta t$ after the passage of the wave, the width of the ribbon zone is $V\Delta t$. As a consequence, in spite of the clotting mechanism, one may wonder whether such a ribbon of fluid *WL*, mechanically equivalent to a shear crack, may become unstable or remain stable. This is a "Griffith-like" problem, but its specific geometry requires a particular treatment. We show in the Appendix that, whatever the ribbon width, the system is always stable. The clotted zone is thus expected to eventually expand on the whole slope, probably associated with an audible "whumpf", but without any avalanche release.

We shall illustrate now these two cases on a numerical example close to typical field conditions, comparing applied and critical stresses for different slope angles. We found in section 2 that the fluid/solid transition takes place for a critical $\Sigma$ value $\Sigma^* \approx 7.04$. However, as we do not have numerical figures for the proportionality coefficient before the exponential in Eq. (5) and for the pure "liquid" viscosity $\eta_o$ involved in Eq. (11), nor any values from specifically dedicated experiments, we shall make here very crude estimates from the literature, in order to obtain nothing but an order of magnitude for the critical shear stress $\tau^*$.

The critical shear stress $\tau^*$ has been indirectly estimated from crack face friction experiments (Rutsch Block or PST) [19]. In some of their experiments, the slab comes quickly to a rest (e.g. in their Fig. 4) instead of accelerating downslope (in their Fig. 2), which is a clear illustration of a very sharp *F/S* transition. Indeed, their experiments *C5* and *C1* for instance exhibit resp. an accelerating and a clotting behaviour, both for a slope angle of 33º, suggesting that both observations are made very close to the transition. With α≈33º and a slab depth around h≈30 cm (see their Fig. 1), and taking a slab density of 300 kg/m$^3$, the transition should occur for a critical shear stress $\tau^* \approx 480$ Pa.

Such a critical stress has to be compared with the shear stress experienced by the *WL*:

$$\tau = \rho g h \sin \alpha = \rho g h_{//} \cos \alpha \sin \alpha = \frac{\rho g h_{//}}{2} \sin 2\alpha \qquad (13)$$

This is illustrated in Fig. 3, for different slab depths ranging from 10 to 40 cm. The *sin2α* function responsible for the inverted U-shaped shear stress curves results from our assumption that snow falls vertically in average, giving a constant vertical h$_{//}$ depth. It goes to zero for slope angles of both zero and 90$^o$, since on moderate slopes the actual slab weight is large but its shear component tends to zero, whereas on steepest slopes the shear component is large, but the amount of snow tends to zero. The horizontal line shows the critical stress estimate $\tau_c$ = *480 Pa* for this particular example. It intersects the shear stress experienced by the *WL* for slab depths larger than about 33 cm. For a slab depth of 40 cm, avalanches can be triggered for slope angles larger than 28º, and up to 62º.

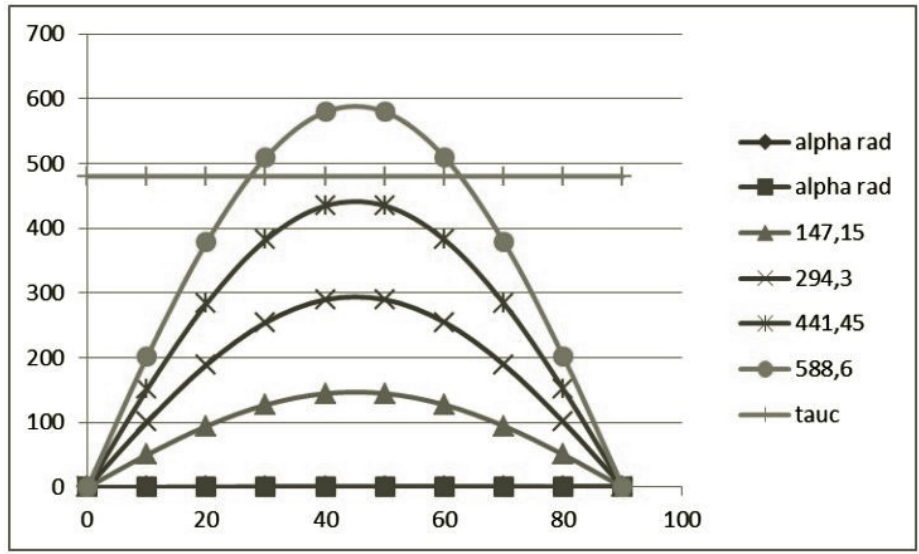

**Fig. 3: Comparison of the critical shear stress τ\* with the shear stress τ experienced by the weak layer:** *the taken slab density is 300kg/m³. Slab depths (measured vertically) are 10cm (triangles), 20cm (crosses), 30cm (stars) and 40cm (disks). In this example, avalanche release cannot occur for slab depths equal to or less than 30cm. For a depth of 40 cm, avalanching becomes possible above a slope angle of 28º, and up to a slope angle of 62º (see text).*

From Fig. 3, and within the limits of our assumptions, inferences are as follows:

i) low slopes: large normal-slope stresses result in *WL* crushing (mainly by collapse) and associated whumpfs; slope-parallel motion quickly comes to a rest due to clotting of the fluid phase, and avalanches are not released.

ii) intermediate slopes: The shear stresses and the associated shear strain rates are sufficiently large to keep the *WL* in the fluid state, and the avalanche is likely to be released.

iii) steep slopes: same conclusions as in i). Such a prediction may seem to be surprising at first glance, but, due to shallow slabs, shear stresses may be reduced below the critical value τ\*. Crown cracks are indeed sometimes observed to open at the junction between intermediate and very steep slopes.

iv) The question arises whether, after clotting (i.e. whumpfing without avalanche release, as in i)), the clotted *WL* may be detabilized again by an additional mechanical shock, as mentioned in the introduction, for instance during a skier's sharp turn or fall. However, beyond the fact that the clotted *WL* is significantly tougher than the initial non-collapsed one, as clotting took place for stresses lower than τ\* (bottom curve in Fig. 1), a possible shear crack due to a local solid to fluid transition in the *WL* and experiencing the same stress would not propagate. The slope would remain safe, except possibly after another snowfall that would bring the *ψ(Nc)* curve of Fig. 1 from the bottom curve to the upper one.

## 4. Discussion and concluding remarks

 In agreement with [19], our model predicts a sharp transition between whumpfing and avalanching, but in contrast with this paper, this sharp transition is not discussed here in terms of an evolution of a Coulomb crack-face friction stress between two solids, that should depend on (and increase with) normal stress, but in terms of a sudden change in viscosity during a phase transition from a fluid to a solid phase (or conversely) in the collapsed *WL,* controlled by the shear strain rate. By contrast with Coulomb's case, an load increase (i.e. an increasing shear stress) results here in a <u>decrease</u> of viscosity. The huge viscosity discontinuity between *S* and *F* phases evidenced in field experiments [1] (typically from 10 to about $10^5$ Pa.s) confirms indeed that a friction analysis based on Coulomb friction is questionable in the present case.

The above results stand for the ideal case of <u>infinite</u>, <u>smooth</u> and <u>uniform</u> slopes. By contrast, real terrain may show slope changes and specific boundary conditions that may affect the results.
For instance, a human-triggered collapse may occur on flat terrain (e.g. a thalweg), and propagate to adjacent slopes. If snow depth and slope on the adjacent slope result in a shear stress larger than $\tau^*$, (top curve of Fig. 1), an avalanche would be released from the slope, and possibly bury the people who triggered the collapse in the thalweg.

On the opposite, the shear stress computed for a uniform slope may be reduced by boundary conditions (stauchwall, gully banks, …). Such a decrease (and distortion) of the applied shear stress curve of Fig. 3 shows that the critical angle for avalanche release is expected to increase, and the slope may remain stable when the distorted shear stress curve lies entirely below the critical shear stress $\tau^*$ (horizontal line).

In addition, snow transportation by wind may locally increase snow depth and also result in a distortion of the inverted U-shaped curves, in particular for large slope angles, favoring avalanche triggering on such slopes.

Finally, slab release not only requires downslope slab shift, but also slab rupture, i.e. crown crack opening [7,20]. This last point is scarcely mentioned in slab release models. Crown crack opening takes place when the increasing tensile load experienced by the upper zone of the slab (due to the slope parallel component of the destabilized slab weight) exceeds slab tensile strength, yielding avalanche release. Despite the fact that the fluid ribbon surrounding the clotted disk is stable vs a shear crack expansion, as shown in the Appendix, if its width is large enough, a crown crack may open at its junction with the clotted disk, releasing an avalanche.

In summary, the present model is based on previous field observations [1] showing that crushed weak layers are made of a granular material that can experience sudden phase transitions from a granular fluid phase to a granular solid one, and conversely.

On this basis, we derived evolution equations, considering shear-rate dependent erosion and aggregation kinetics of ice grains, that qualitatively eproduce the observed behavior. The model is then applied to the slab avalanche triggering problem. It predicts a sharp transition between whumpfing and avalanching, controlled by the shear strain rate of the collapsed weak layer. Within our simplifying assumptions, avalanching becomes theoretically possible when the slope angle exceeds a value typically between 10º and 40º (Fig. 3). However, such results, obtained for infinite, smooth and uniform slopes, are likely to be modified on real slopes, where stauchwalls or other boundary constraints may shift the critical slope angle up to higher values, in closer agreement with statistical data, as reported for instance on the "data-avalanche" website [21]. By contrast, snow transportation that may increase slab depth at places is likely to favor avalanche triggering. A local slope increase may also favor a tensile crown crack opening (e.g. at the junction between clotted and still unclotted zones) and destabilize the system.

Owing to its general character, our model may be used in other cases where granular slurries may undergo abrupt viscosity changes upon continuous shear strain rate changes or sudden mechanical loading. In particular, our model may account for the sudden arrest of dense avalanche flows as vanishing slopes reduce the shear strain rate beyond the critical value, or for concrete solidification in too slow spinning mixers. More hypothetically, it may also be relevant for investigating permafrost stability against building construction [22,23].

**Acknowledgements**

The author is indebted to J. Blackford, S. Caffo, A. Duclos, J. Heierli and J. Weiss for helpful comments and discussions.

**References**

1- A. Duclos, S. Caffo, M. Bouissou, J.R. Blackford, F. Louchet, J. Heierli, Granular Phase Transition in Depth Hoar and Facets: A New Approach to Snowpack Failure?, paper presented at the International Snow Science Workshop, Davos, Switzerland, 27 september-2 october 2009. URL: http://www.slf.ch/

2- M.J. Powell, Site percolation in randomly packed spheres. *Physical Review B*. **20**(10), 4194-4198 (1979), doi: 10.1103/PhysRevB.20.4194.

3- R. Consiglio, D.R. Baker, G. Paul, H.E. Stanley, Continuum percolation thresholds for mixtures of spheres of different sizes. *Physica A* **319**, 49-55 (2003).

4- A.J. Hughes, The Einstein relation between relative viscosity and volume concentration of suspensions of spheres. *Nature* **173**, 1089-1090 (1954), doi: 10.1038/1731089a0.

5- G.A. Campbell, G. Forgacs, Viscosity of concentrated suspensions: An approach based on percolation theory. *Physical Review A* **41**(8), 4570-4573 (1990).

6- A.A. Griffith, The phenomena of rupture and flow in solids. *Philosophical Transactions of the Royal Society, London* **221**, 163-198 (1920).

7- F. Louchet, A simple model for dry snow slab avalanche triggering. *Comptes Rendus à l'Académie des Sciences Paris* **330**, 821-827 (2000).

8- F.Louchet, J. Faillettaz, D. Daudon, N. Bédouin, E. Collet, J. Lhuissier, A-M. Portal Possible deviations from Griffith's criterion in shallow slabs, and consequences on slab avalanche release. *Natural Hazards and Earth System Sciences*, **2**(3-4), 1-5. (2002).

9- D. McClung, Shear fracture precipitated by strain softening as a mechanism of dry slab avalanche release. *Journal of Geophysical Research* **84**(B7), 3519-3526 (1979).

10- D.McClung, Fracture mechanical models of dry slab avalanche release. *Journal of Geophysical Research* **86** (B11), 10783-10790 (1981).

11- Z.P. Bazant, G. Zi, D. McClung, Size effect law and fracture mechanics of the triggering of dry snow slab avalanches. *Journal of Geophysical Research* **108**(B2), 2119, 13-1 - 13-6 (2003), doi: 10.1029/2002JB001884.

12- B. Johnson, B., Jamieson, B., and Johnston, C., Field Data and Theory for Human Triggered "Whumpfs" and Remote Avalanches*, International Snow Science Workshop*, Big Sky, MT, USA, Ed. Montana State University, MT, USA, 2 - 6 october 2000, 208-214. URL: http://www.issw.net/2000.php

13- J.B. Jamieson, J. Schweizer, Texture and strength changes of buried surface-hoar layers with implications for dry snow-slab avalanche release. *Journal of Glaciology* 46(152), 151-160 (2000). doi: http://dx.doi.org/10.3189/172756500781833278.

14- D.Gauthier, B. Jamieson, Evaluation of a prototype field test for fracture and failure propagation propensity in weak snowpack layers. *Cold Regions Science and Technology* **51**(2-3), 87-97 (2006).

15- D. Gauthier, B. Jamieson, Fracture propagation propensity in relation to snow slab avalanche release: Validating the Propagation Saw Test. *Geophysical Research Letters* **35**, L13501-13504 (2008). doi: 10.1029/2008GL034245.

16- J. Heierli, A. Van Herwijnen, P. Gumbsch, M. Zaiser, Anticracks: a new theory of fracture initiation and fracture propagation in snow, *International Snow Science Workshop*, Whistler, BC, Canada, Ed. Montana State University, MT, USA, 21-27 september 2008, 9-15. URL: http://issw.net/2008.php

17- J. Heierli, P. Gumbsch, M. Zaiser, Anticrack Nucleation as Triggering Mechanism for Snow Slab Avalanches. *Science* **321**, 5886, 240-243 (2008). doi: 10.1126/science.1153948.

18- J. Heierli, Solitary fracture waves in metastable snow stratifications, *Journal of Geophysical Research* **110**, F02008, 1-7, (2005) doi: 10.1029/2004JF000178.

19- A. Van Herwijnen, J. Heierli, Measurement of crack-face friction in collapsed weak layers, *Geophysical Research Letters* **36**, L23502, 1-5 (2009)

20- J.Faillettaz, F. Louchet, J-R. Grasso, Two-threshold model for scaling laws of non-interacting snow avalanches. *Physical Review Letters* **93** (20), 208001-1 - 208001-4 (2004). doi: 10.1103/PhysRevLett93.208001.

21- Data-avalanche (http://www.data-avalanche.org/)

22- J.Vakili, Slope stability problems in open pit coal mines in permafrost regions, paper presented at the International Arctic Technology Conference, Anchorage, Alaska, USA,29-31 may 1991, AIME, Society of Petroleum Engineers (1991). doi: 10.2118/22141-MS. URL: http://www.amazon.com/International-Arctic-Technology-Conference-Proceedings/dp/9991136738

23- F.E. Crory, Piling in frozen ground, *J. of the Technical Councils of the American Society of Civil Engineers*, **108**(1), 112-124 (1982).

**Appendix**

**Stability of a clotted WL disk surrounded by a ring-shaped fluid WL zone**

The still uncollapsed ribbon surrounding the clotted central disk of the WL is considered as a shear crack, as it opposes a very small resistance for slide as compared with that of the central clotted disk. The crack stability in such a configuration is a Griffith-like problem, but its specific geometry requires a particular calculation of the critical radius.

We consider a collapsed zone of radius $\Lambda = Vt$, where $V$ is the velocity of the solitary wave bounding the collapsed zone, and $t$ the time elapsed since collapse initiation at point O. As mentioned in the main text, after a time $\Delta t$, the central area starts clotting. If the clotted zone has a radius $r$, the still unclotted ring between the wave front and the clotted zone has a width $\Lambda$-$r$.

As in the classical treatment of the Griffith problem, we compare the variations of the stored elastic energy and the energy required for expansion of the unclotted zone equivalent to a crack (crack opening energy), experiencing a shear stress $\tau$.

Let us consider the torus centered in O, with an external radius $\Lambda$ and an inner radius $r$, embedding the unclotted area. The radius of the circle centered on O and running in the middle of the ring is *($\Lambda$+r)/2*, and the radius of the torus section is *($\Lambda$-r)/2*. Therefore, the volume of the torus centered on O, that embeds the unclotted ring, is:

$$\Omega = 2\pi^2 \left(\frac{\Lambda - r}{2}\right)^2 \left(\frac{\Lambda + r}{2}\right) = \frac{\pi^2}{4}(\Lambda - r)^2(\Lambda + r) \tag{A1}$$

and its variation for a fluctuation $dr$ of the clotted zone radius $r$ is given by:

$$\frac{d\Omega}{dr} = \frac{\pi^2}{4}\left(3r^2 - 2\Lambda r - \Lambda^2\right) \tag{A2}$$

At a time $t$, i.e. for a given $\Lambda$ value, the energy balance between the variations of the stored elastic energy $W_1$ and the crack opening energy $W_2$ for a fluctuation $dr$ of the crack size around the clotted zone is:

$$\frac{dW}{dr} = \frac{d(W_1 + W_2)}{dr} = \frac{\tau^2}{2E}\frac{d\Omega}{dr} - 2\pi\gamma\, r = \frac{\tau^2}{2E}\frac{\pi^2}{4}\left(3r^2 - 2\Lambda r - \Lambda^2\right) - 2\pi\gamma\, r \tag{A3}$$

where $\tau$ is the shear stress, $\tau^2/2E$ the corresponding stored elastic energy density, and $\gamma$ the free surface energy.

From eq. (A3), the critical "Griffith" radius $r*$ is solution of the equation:

$$\frac{\pi^2\tau^2}{8E}\left(3r^2 - 2\Lambda r - \Lambda^2\right) - 2\pi\gamma\, r = 0 \tag{A4}$$

This $2^{\text{d}}$ degree equation obviously has one positive and one negative roots. The positive root $r*$ is given by:

$$r* = \frac{1}{3}\left(\Lambda + \frac{8E\gamma}{\pi\tau^2} + \sqrt{4\Lambda^2 + 2\Lambda\frac{8E\gamma}{\pi\tau^2} + \left(\frac{8E\gamma}{\pi\tau^2}\right)^2}\right) \tag{A5}$$

In addition, the coefficient of $r^2$ in eq. (A4) being positive, $dW_1 > dW_2$ for $r > r*$. The ring-shaped crack is therefore unstable for $r > r*$.

However, the critical radius $r*$ given by eq (A5) is obviously larger than:

$$r_0* = \frac{1}{3}\left(\Lambda + \sqrt{4\Lambda^2}\right) = \Lambda \tag{A6}$$

As $r$ is always smaller than $\Lambda$, the ring-shaped crack is never unstable.

It can be noticed that if the radius of the torus section $(\Lambda - r)/2$ is large as compared to slab depth, the elastic energy would be mainly stored in the lower part of the torus (older snow). This modification would change $8E\gamma$ into $16E\gamma$ in eq. (A5), but eq. (A6) would remain unchanged, as well as the conclusions.